**Utsarjan: A smartphone App for providing kidney care and real-time assistance to children with nephrotic syndrome**

Snigdha Tiwari[1], Sahil Sharma[1], Arvind Bagga[2], Aditi Sinha[2], Deepak Sharma[1,*]

[1]Computational Biology and Translational Bioinformatics (CBTB) Laboratory, Department of Biosciences and Bioengineering, Indian Institute of Technology Roorkee, Roorkee 247667, Uttarakhand, India.

[2]Department of Pediatrics, All India Institute of Medical Sciences, New Delhi 110029, India.

*Corresponding Author

Deepak Sharma, Department of Biosciences and Bioengineering, Indian Institute of Technology Roorkee, Roorkee 247667, India; Tel: +91-1332-284827; Emails: deepak.sharma@bt.iitr.ac.in; deepak.aiims@gmail.com

# GRAPHICAL ABSTRACT

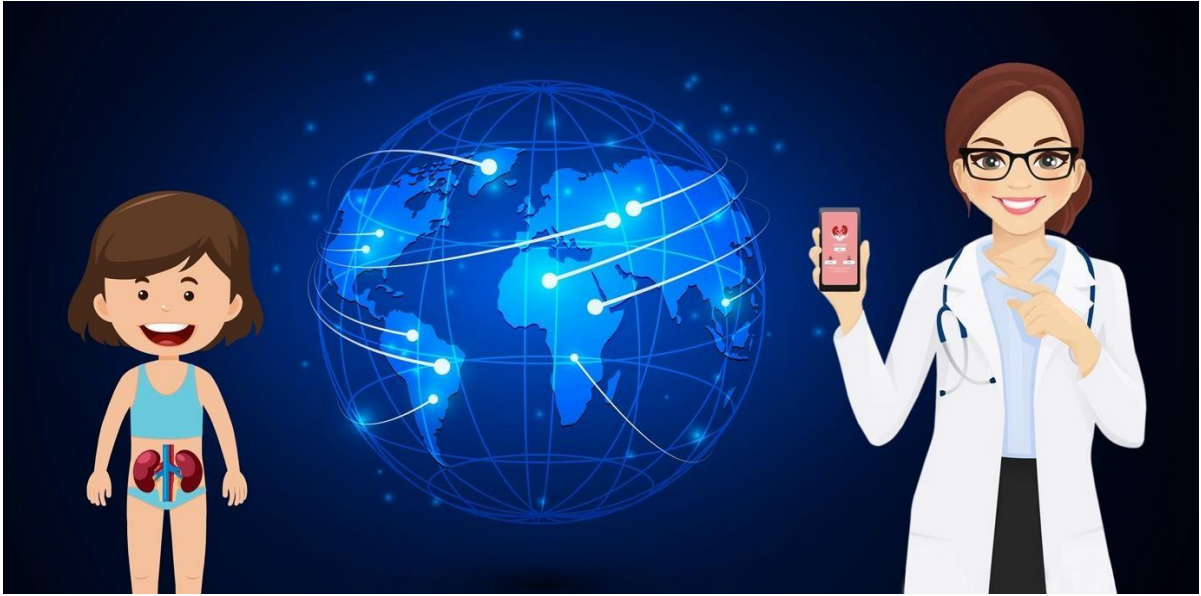

# ABSTRACT

**Background**

Telemedicine has the potential to provide secure and cost-effective healthcare at the touch of a button. Nephrotic syndrome is a chronic childhood illness involving frequent relapses and demands long/complex treatment. Hence, developing a remote means of doctor-patient interface will ensure the provision of quality healthcare to patients.

**Methods**

The Utsarjan mobile App framework was built with Flutter that enables cross-platform development (Android, iOS, Windows) with speed, smoothness, and open-source benefits. The frontend uses Dart for user interaction, while the backend employs Node.js, Express, and NGINX for APIs, load balancing and high performance. MongoDB ensures a flexible database, Bcrypt secures passwords, PM2 handles deployment, uptime and logs, while Firebase Cloud Messaging powers free push notifications.

**Results**

Utsarjan (means excretion) is a multi-functional smartphone application for giving nephrotic care and real-time assistance to all patients (especially those in rural regions and/or who do not have access to specialists). It helps patients and doctors by ensuring opportune visits, recording each clinical test/parameter and improving medication adherence. It gives a graphical visualization of relapses, medicine dosage as well as different anthropometric parameters (urine protein, BP, height and weight). This is the first nephrotic care App that enables prompt access to doctor's advice.

**Conclusions**

Utsarjan is a mobile App to provide kidney care and real-time assistance to children with nephrotic syndrome. It gives a graphical overview of changes in a patient's health over the long course of treatment. This will assist doctors in appropriately modifying the treatment regimen. Consequently, it will (hopefully) lead to the prevention of relapses and/or complications.

## 1. Introduction

Nephrotic syndrome is a chronic childhood illness characterised by the tetrad of proteinuria, hypoalbuminaemia, hyperlipidaemia and oedema [1]. The most common type of nephrotic syndrome is primary (idiopathic) with an incidence of 1.15-16.9 per 100000 children [2]. Notably, the incidence is quite high (~16.0) in South Asian countries. On the basis of response to steroid therapy, nephrotic syndrome is categorized as Steroid-sensitive (SS) and Steroid-resistance (SR). Steroid-responsiveness and the frequency of relapses determine the course of treatment. Management of nephrotic syndrome includes specific as well as supportive treatments. Idiopathic nephrotic syndrome encompasses a relapsing-remitting course within the large share of pediatric patients, requiring a careful observation and self-management. Children with steroid resistant diseases are at increased risk of progression to end-stage renal disease (ESRD) [3].

Nephrotic syndrome requires careful medical supervision, frequent hospital visits and active parental involvement in management. Hence, a mobile App is expected to reduce delays in therapy, improve therapy adherence, address parental concerns, improve quality of life and by reducing hospital visits, decrease healthcare costs and school absenteeism [4]. Furthermore, in situations like the recent COVID-19 pandemic, hospitals have constrained access to in-person visits to avoid infection transmission [5]. Medicoinformatics has become crucial for delivering quality healthcare, with its adoption accelerating post-COVID-19, transforming medical care globally [6,7]. In spite of the fact that there are numerous kidney-related Apps, nearly all of these are restricted to providing disease-related data and lack the association of clinicians. Several studies have proposed the requirement of a nephrotic syndrome App that (i) gives instantaneous access to doctor's advice, (ii) empowers two-way communication between patients and doctors for real-time restorative help and has shown to be clinically effective [8,9]. An App that incorporates such features will be of great utility.

Towards this end, we have developed a multi-functional, interactive smartphone application, Utsarjan (means excretion), for providing care and real-time medical support to nephrotic syndrome patients [10]. The App acts as a doctor-patient interface to help the patient to record their health periodically and get instantaneous access to the doctor's advice. The App will bridge geographical barriers by enabling virtual visits (instead of in-person visits). Consequently, it will lead to improvement in the follow-up of the patients. The App will also be useful in educating patients (and their family members) about how to take the best care of kidney health. As children are often dependent on their parents/caregivers for care, the knowledge and involvement of caregivers can significantly impact the management of the condition. In the long run, the development of this platform will lead to the collection of information (incidence, prevalence, age at onset and demographic data) about the different categories of patients.

## 2. Methodology

The Utsarjan mobile App framework is built upon robust, production-grade backend technologies and agile software development lifecycles, enabling it to effectively manage distributed systems with high speed and throughput. The App has been developed using the Flutter platform that not only allows building Apps for Android, iOS and Windows users but also offers enhanced speed, smoothness, ease of development, and the added benefit of being free and open-source software (Fig. 1). The App has been built using the Flutter architectural layers, leveraging fundamental classes for essential functionalities, rendering layers to handle layout, and the widgets layer for creating a hierarchical composition of elements.

The frontend, written in Dart, facilitates user interaction. Various Flutter packages enhance the App's functionality by providing key features, ensuring a seamless and robust user experience. This framework includes essential tools and components to design dynamic and efficient UIs [11]. The 'flutter_localizations' package provides essential multi-language support for the App, enabling seamless localization to cater to various regions and cultures. By integrating this package, the App can dynamically adjust its content, labels, and formatting to suit users from different linguistic backgrounds, enhancing the overall user experience for a global audience. The 'font_awesome_flutter' package brings a vast collection of pre-designed icons from the renowned Font Awesome library, offering developers an easy and efficient way to integrate high-quality, functional icons into their App's interface. This not only enhances the visual appeal but also contributes to intuitive navigation and a richer user experience, with minimal effort required for customization. The 'flutter_local_notifications' package provides functionality for displaying local notifications within the App, making it particularly useful for creating alerts, reminders, or prompts that engage users and keep them informed of important events or actions. This capability is ideal for reminding users of scheduled tasks, deadlines, or App-specific updates without requiring an active internet connection. The 'scoped_model' provides a state management solution for Flutter, allowing easy sharing of data and state across different parts of the App by using models. The 'url_launcher' package simplifies interaction with external systems and Apps by enabling actions such as opening web URLs, dialing phone numbers, sending emails, or SMSes directly from within the App. This package helps create a more integrated experience, allowing users to perform these actions without needing to leave the App, enhancing convenience and usability. 'Permissions_Security' package is crucial for managing and requesting permissions, such as accessing the camera, location, contacts, or other sensitive areas of the user's device. The 'pinch_zoom' and 'photo_view' packages both enhance user interaction with images by enabling pinch-to-zoom functionality. This feature allows users to easily zoom in and out on images with natural gestures, improving the usability of Apps that involve detailed visuals, such as galleries, maps or documents. These packages provide smooth and intuitive zooming capabilities, making it easier for users to explore finer details within images. On the multimedia front, 'youtube_player_flutter' enables seamless embedding and playback of YouTube videos directly within the App. This integration allows users to view video content without leaving the App, offering a richer, more immersive multimedia experience.

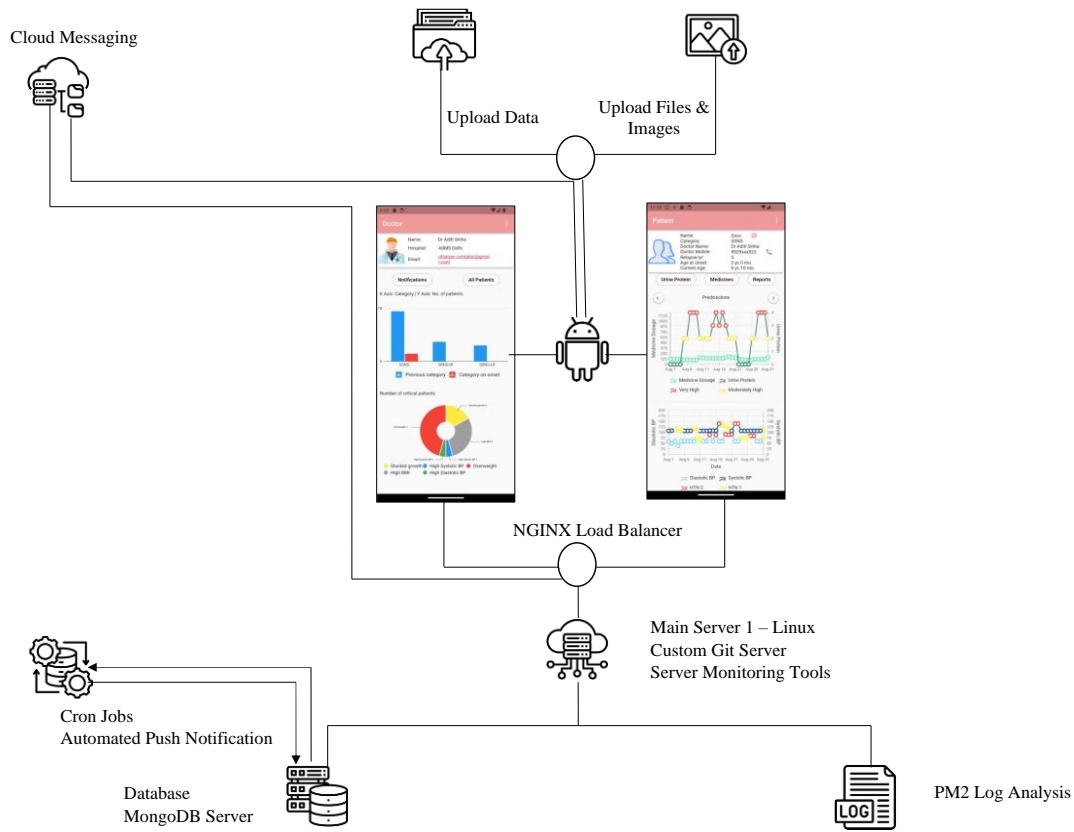

**Fig. 1:** System design of Utsarjan mobile App.

The backend APIs for this project were developed using the Node.js framework and hosted on an Ubuntu server, providing robust and scalable data management [12]. Several essential Node.js packages were employed to enhance functionality and streamline operations. 'Mongoose (MongoDB)' package establishes a connection between the MongoDB database and the server, allowing for the creation, validation, and manipulation of schema-based models, ensuring smooth data handling and structured storage. 'Express' framework was used for building RESTful APIs; it simplifies the creation of endpoints and routes, making it easier to manage the API logic within the Node.js environment. 'CORS (Cross-Origin Resource Sharing)' was implemented to manage and enable secure cross-origin access to the APIs, ensuring that requests from different domains are handled appropriately and securely. 'Moment' utility was used for handling date-related functionality, simplifying operations involving date and time formatting, manipulation, and validation. 'Morgan' acting as a middleware, was used to log HTTP requests and errors, providing valuable insights into server performance and assisting in debugging issues. 'Multer' was employed for file uploading, particularly to handle user-related media such as profile photos, ensuring efficient file storage and management on the server. 'Nodemailer' package allows the application to send formatted emails; in our App it was used for sending OTPs (One-Time Passwords) to users' email addresses as part of the authentication and verification process. 'FCM-Node' was integrated for push notification functionality; it sends real-time alerts and notifications directly to users' devices, improving user engagement. 'Body-Parser' parses incoming HTTP request bodies, enabling the server to properly handle and interpret incoming data, such as JSON or form submissions. 'Async-CSV' tool was used to parse and manipulate CSV data, allowing for efficient handling of CSV files, which can be vital for tasks like data import/export.

To ensure high performance and load balancing, the client sends requests to the server with NGINX serving as the HTTP server, reverse proxy and load balancer. The code follows a highly modular microservices architecture, promoting the continuous delivery/deployment of large, complex applications, and can be reused easily. It uses the recycler views and highly efficient coding, for the purpose of maintaining memory efficiency and reducing application size. The 'Bcrypt' library provides a password-hashing function utilized for securely hashing passwords, transforming them into alphanumeric representation. The App's deployment and management in production are facilitated by the use of the PM2 toolbox ensuring uptime, managing processes, and providing access to logs/clusters. Additionally,

Firebase service like Cloud Messaging has been integrated into Utsarjan App to reliably send push notifications to doctors and patients at no cost.

**3. Results**

Utsarjan is designed to assist children with nephrotic syndrome and their caregivers by addressing key challenges. The App helps in maintaining records of clinical parameters, symptoms, tests, and relapses while also tracking visits and medication with timely reminders. It sends automatic alerts to doctors and patients if any parameter exceeds the normal range. It ensures wide usability by supporting multiple languages (like English and Hindi). Most importantly, it facilitates real-time two-way communication between patients and doctors for immediate medical support.

*3.1. Features for patients in Utsarjan App*

Patients have the freedom to register by clicking on the 'Patient' button displayed on the home screen with/without mentioning the doctor's details (Fig. 2a). After login, patients are directed to their personalized home screen (Figs. 2b and 2c), which showcases their profile information and provides access to the following features:

*3.1.1. Graphical/textual visualization*

The App provides a comprehensive data visualization through graphs, allowing patients to easily track their health status. The graph displays information on medicine dosage and urine protein (Fig. 2b), systolic/diastolic BP (Fig. 2b), body height/weight (Fig. 2c) and BMI (Fig. 2c) for an easy overview of all the data uploaded to date. To enhance user understanding, the data points on the graph are color-coded for easy interpretation (moderate fluctuations are coloured yellow while higher changes are coloured red). For instance, slightly higher urine protein levels (30+/100++) are denoted with yellow points. In contrast, red points indicate very high urine protein (300+++/2000++++) that leads to relapse if consistently observed three times in a row. Likewise, slightly elevated BP (Stage1 HTN) is marked with a yellow point while high BP (Stage2 HTN) with a red point [13]. Height, weight and BMI data for girls and boys are compared to Median and Standard Deviation (SD) values; the data points are coloured red for values ≥2|SD| and yellow for values between <2|SD| to ≥1|SD| [14]. In case of alarming/critical situation, the 'healthy kidney' sign is replaced by 'fire/warning' sign. Furthermore, beneath the graphs, complete details of the patient's data is shown enabling easy reference and accessibility whenever needed (Fig. 2l).

### 3.1.2. Adding data

Through the 'Urine Protein' button available in the App (Fig. 2d), patients have the option to input their urine protein levels along with any additional symptoms or comments relevant to their clinical data. A crucial aspect of the App is its proactive alert system. Whenever any parameter goes beyond the normal range, the App automatically triggers a notification to alert the patient (Fig. 2f). Simultaneously, the doctor also receives the notification that the particular patient's condition requires immediate attention (Fig. 2j). This timely alert system ensures that both patients and doctors are promptly informed of any concerning developments, allowing for swift intervention and proactive management of the patient's health.

### 3.1.3. Sending clinical reports

The App incorporates another valuable feature that enables patients to seamlessly send images from their device's storage library using the 'Reports' button to their respective doctors (Fig. 2b). By analysing these reports, doctors can make informed decisions and advise appropriate clinical tests and medications based on the patient's health condition.

### 3.1.4. Medication profile

The 'Medicines' button serves as a tool for patients to keep a track of all the prescribed medications (Fig. 2e). Patients can update the status of each medication, indicating whether it has been taken or not. These records are then utilized to generate a dosage graph on the main page, providing patients with a visual overview of their medication intake (Fig. 2b). The App empowers patients to manage their medications effectively, fostering adherence to treatment plans and promoting better health outcomes.

### 3.1.5. Additional useful features

The 'Drop down' menu on the top right corner of the patient's home screen provides several useful features to enhance patient experience and facilitate management of their healthcare needs including (i) 'How To Use App', (ii) how to record urine protein by dipstick; and (iii) information about nephrotic syndrome, including its symptoms, complications, options of management, parental responsibility, dietary advice, advice on vaccination and long-term outcomes, (iv) 'Notifications', wherein all the notifications sent by the doctor, clinical tests advised, medicine updates and relapse notifications are listed, (v) 'Nearby Hospitals', displays a map of all the hospitals that can be accessed in a short time in case of any emergency, (vi) 'Change/Add Doctor', gives freedom to the patient to change a doctor/center and all the data automatically transfers to the new doctor or add a doctor, if the patient is presently not registered with any doctor and (vii) 'About the Disease', gives exhaustive information of the disease through FAQs about Nephrosis (in Hindi and English languages) and videos.

## *3.2. Features for Doctors in Utsarjan App*

Like patients, doctors also need to first register themselves using the 'Doctor' button on the home screen of the App (Fig. 2a). Subsequently, they can log in by clicking the 'Sign In' button to access the following features:

### *3.2.1. Overview of all patients*

Upon signing in, the doctor gains access to a graphical representation of the categories of all the patients [Steroid Sensitive Nephrotic Syndrome (SSNS), Steroid Resistance Nephrotic Syndrome-Initial Resistance (SRNS-IR) and Steroid Resistance Nephrotic Syndrome-Late Resistance (SRNS-LR)] (Fig. 2g). The doctor's home screen also features a doughnut chart that provides an overview of the number of patients that are critical [15].

### *3.2.2. Access to patient data*

A doctor can view a comprehensive list of all the registered patients by clicking the 'All Patients' button on the home screen (Figs. 2g and 2h). Most importantly, the App automatically sends notifications in case of relapse. The 'All Notifications' button on the home screen enables the doctor to see all the notifications sent to patients (Figs. 2g and 2j). For each patient, the doctor can access detailed information, including a graphical representation of medicine dosage, urine protein, systolic/diastolic BP, height, weight, and BMI (Fig. 2i). If needed, doctor can also download all the data. Furthermore, the doctor has the option to authenticate his/her patients (indicated by a 'tick') [Fig. 2i].

### *3.2.3. Adding Data*

The App provides doctors with the option to input various parameters of a patient such as systolic/diastolic BP, body height/weight, BMI (automatically calculated), onset category and any other extra comments/complications (Fig. 2k). The data entered is represented in a graphical form which allows the doctor to observe and analyze the patient's health and progress over time.

### 3.2.4. Prescribing medicine

The doctor can view the medicines prescribed to any patient and can update them if required (Fig. 2i). The doctor has the freedom to modify/add the 'Steroids' and 'Other Medicines' categories.

### 3.2.5. Notify patients

If required, the doctor can send advice/notification to a particular patient through the 'Notify' button (Figs. 3a and 3b). Additionally, the App provides a call functionality that allows the doctor to contact the patient directly from within the App (Fig. 3a).

### 3.2.6. Register new patients

Doctors also have the freedom to register new patients if required, enabling them to maintain a comprehensive record of each patient's medical history. During each visit, the doctor can personally update/track their clinical tests and parameters, ensuring accurate and up-to-date data.

### 3.2.7. Additional useful features

On the detailed patient's page, several useful features (Fig. 3a inset) have been incorporated for the doctors: (i) 'Reports' include all the documents that have been uploaded by the patient, (ii) 'Advice/Complaint' feature allows doctors to send appropriate advice to the patient (Fig. 3c); (iii) 'Tests' enable the doctor to send advice for getting the necessary tests done along with the any additional comments (if needed) to the patient (Fig. 3d) and 'Details' page contains all the information related to patient's history which can be updated anytime (Fig. 3e).

### 3.3 Utility of Utsarjan App as digital diary

Parents of children with nephrotic syndrome usually maintain a diary to record information on first morning urine dipstick proteinuria, doses of medications, and symptoms [16]. Utsarjan App helps parents and physicians maintain an electronic diary that is entered on a user-friendly page and can be downloaded as an excel (Figs. 2d, 2e, 2k, 2l and 2m). Treating physicians receive alerts each time a linked patient records urine protein 3+/4+, indicating possible relapse that requires therapy (Fig. 2j). Physicians can review images and reports uploaded by parents and send responses, prescribe medications, and download patient electronic diaries (Figs. 2l and 2m). Each patient's disease course can be monitored comprehensively, including time trends of anthropometry, blood pressure, medication doses and adherence, relapses or proteinuria, and investigations, which aids informed decision-making (Figs. 2g, 2i, 2k, 2l, 2m and 3a).

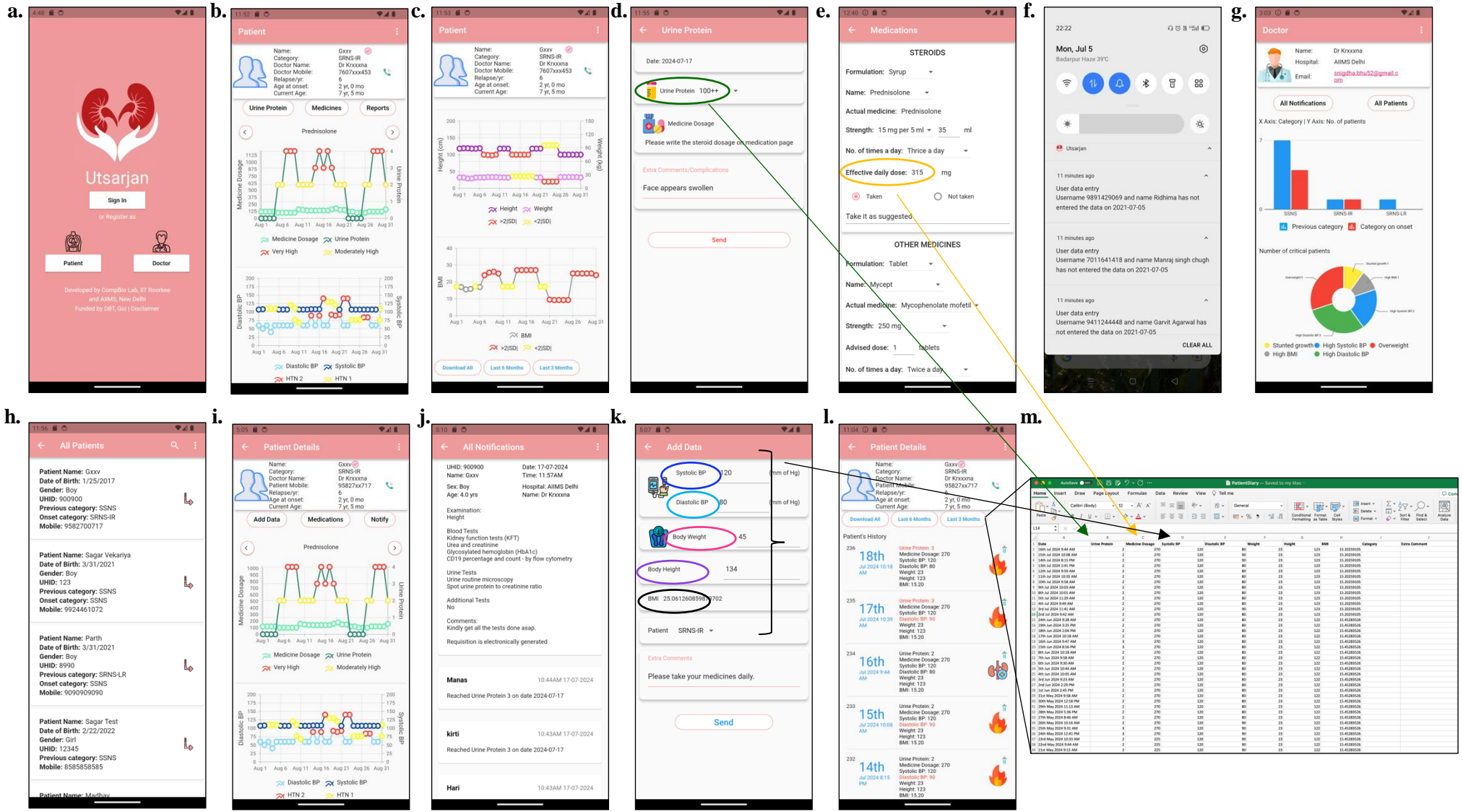

**Fig. 2:** Mobile screenshots of Utsarjan App (a) Homescreen, (b) Patient's profile showing graphical representation of digital diary, (c) Patient's profile showing graphical representation of digital diary, (d) Patient's Urine Protein entry page, (e) Patient's Medications page, (f) Proactive alerts to patients, (g) Physician's homepage, (h) Physician's All Patients page, (i) Patient's profile for physician's monitoring (showing graphical representation of digital diary), (j) Physician's Notifications, (k) Physician's Add Data page, (l) Patient's Details page and (m) Downloaded Excel sheet of Sample digital diary that collates various important patient data [All the names/data used are dummy].

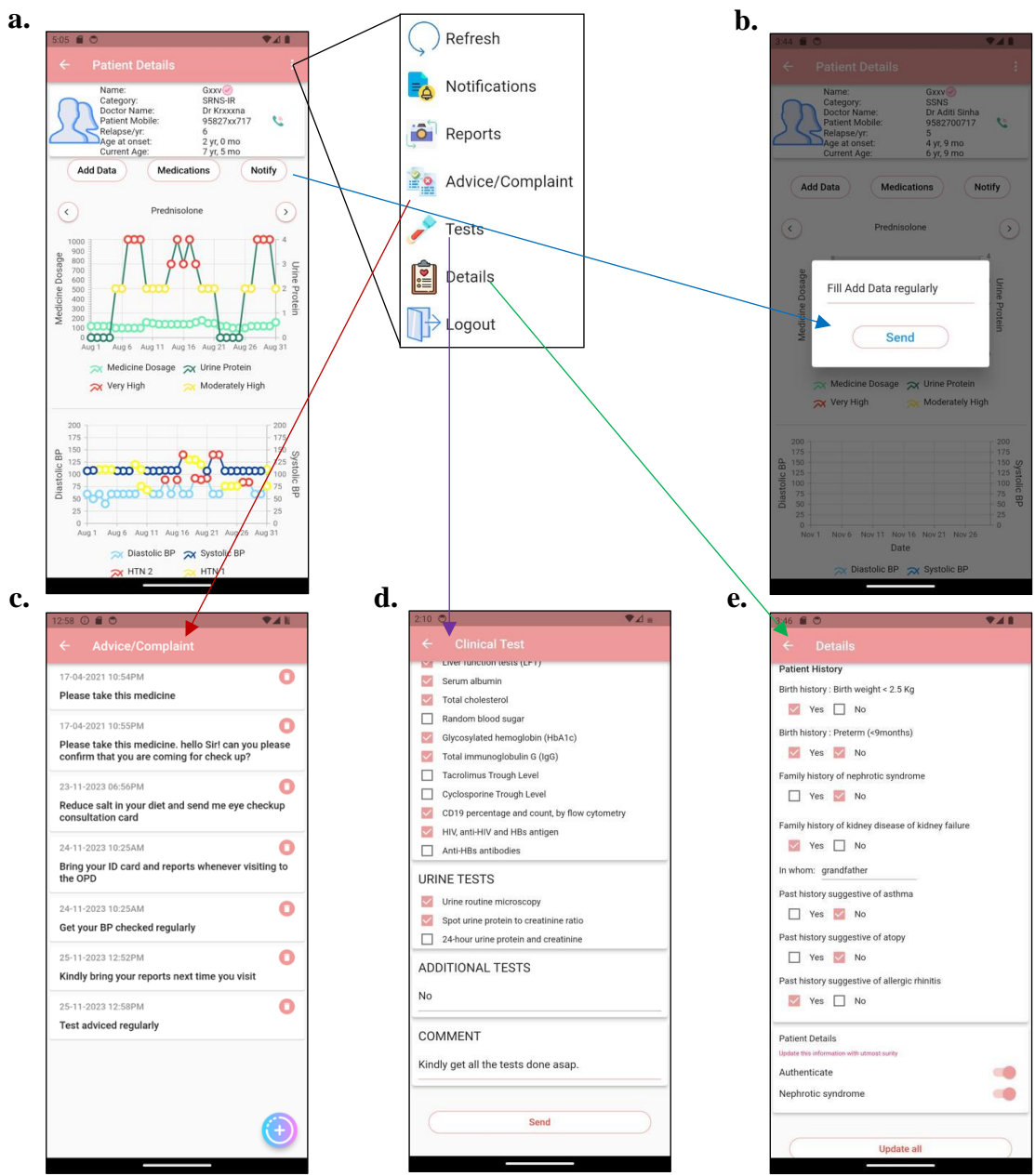

**Fig. 3:** Mobile screenshots of Utsarjan App (a) Patient's profile for physician's monitoring, (b) Notification can be sent by the doctor to patients, (c) Advice/Complaint that patient/doctor can register, (d) Clinical Tests sent by the doctor to patients and (e) Patients detailed page.

## 4. Discussion

Utsarjan is a mobile App designed to provide effective kidney care and real-time assistance to children with nephrotic syndrome. By offering a graphical overview of a patient's health changes during treatment, the App aids doctors in modifying treatment regimens to prevent complications and relapses [17]. The App assists patients and doctors to improve disease tracking, management and treatment. Utsarjan App plays an important role as a digital diary and real- time monitoring can be done by health professionals. The App will improve the interaction between the pediatrician and caregivers in terms of immediate reporting of symptoms, early identification of relapse and providing an opportunity for the pediatrician to manage the condition of the child timely and appropriately. Caregivers are responsible for monitoring andrecording urine testing reports, administering medications, and observing any health issues in the child. A significant proportion of caregivers lack sufficient understanding of home care management for nephrotic syndrome [18]. The inclusion of videos in the Apps serves to impart knowledge to the caregivers [19]. Utsarjan App has 6 videos on demonstration of urine protein monitoring, an introduction to nephrotic syndrome, diet, treatment, immunization, and complications and warning signs regarding nephrotic syndrome.

The 'UrApp' is an existing mobile App that allows remote detection/management of nephrotic relapses [20]. The mHealth App utilizes text-messaging based reminders to improve patient engagement/compliance to medications and urine motoring [21]. Compared to other Apps, the Utsarjan App offers several other features: *(i)* facilitates two-way communications between patients and physicians; *(ii)* growth and blood pressure monitoring; *(iii)* enhance caregivers' knowledge about home-based management using educational videos; *(iv)* informs parents about nearby hospitals and allows them to switch between treating physicians or centers; *(v)* prompt communication of investigations; *(vi)* digital diary; and *(vii)* notifications that enable timely therapy.

The features embedded within the Utsarjan App, such as urine protein monitoring, medication compliance tracking, report uploading, as well as reminders and notifications, that enable timely initiation of therapy, reduce parental anxiety and improve patient adherence to treatment plans and will facilitate early detection of relapses and strengthened interaction between caregivers and healthcare professionals [22]. This is the first pediatric nephrotic syndrome App that provides instantaneous access to doctor's advice. Thus, the Utsarjan App will help in better management of pediatric nephrotic syndrome, potentially reducing complications and improving outcomes. Prospective clinical studies should be carried out to validate the efficacy of Utsarjan App in improving adherence and parental knowledge, and in reducing time to therapy, rates of complications and costs of healthcare. App-based care is readily accepted by the current text-savvy generation of care providers and has potential to complement in-person visits and teleconsultations.

*Availability:* Utsarjan App is freely available on Google Play Store.

**Funding Statement**

DS acknowledges the financial support from MHRD (BT/2014-15/Plan/P-955 and BIO/FIG/100700), SERB (ECR/2016/001566), DBT (BT/PR40141/BTIS/137/16/2021) and DHR (R.11013/51/2021-GIA/HR and R.11017/36/2023-GIA/HR). ST is thankful to MHRD for the research fellowship.

**Competing Interest Statement**

The authors report no conflict of interest.